\documentstyle [11pt] {article}
\makeatletter
\newcommand{\doublespacing}{\let\CS=\@currsize\renewcommand{\baselinestretch}{1.4}\tiny\CS}
\newtheorem{Proposition}{\bf {Proposition}}[section]
\newtheorem{definition}[Proposition]{\bf {Definition}}
\newtheorem{theorem}[Proposition]{\bf {Theorem}}
\newtheorem{example}[Proposition]{\bf {Example}}

\newtheorem{lemma}[Proposition]{\bf {Lemma}}

\begin{document}
\title {\Large {\bf
Convergence of maxgeneralized mean-mingeneralized mean powers of intuitionistic fuzzy matrices}}
\date{}
\author{{\bf  Rajkumar Pradhan  and\,\ Madhumangal Pal
}  \\
 Department of Applied Mathematics with Oceanology and Computer Programming,\\
Vidyasagar University, Midnapore -- 721 102, India.\\
 e-mail: math.rajkumar@gmail.com; mmpalvu@gmail.com}

\maketitle

\subsection* {\centering Abstract }
\begin {quote}
 {\em
 Intuitionistic fuzzy relations on finite universes can be represent by intuitionistic fuzzy matrices and the
 limiting behavior of the power matrices depends on the algebraic operation employed on the matrices. In this
 paper, the power of intuitionistic fuzzy matrices with maxgeneralized mean-mingeneralized mean operation have
 been studied. Here it is shown that the power of intuitionistic fuzzy matrices with the said operations are
 always convergent. The convergence of powers for an intuitionistic fuzzy matrix with convex combination of
 max-min and maxarithmetic mean-minarithmetic mean are also dicussed here.}\\
{\bf \bf Keywords:} {\it Intuitionistic fuzzy number, intuitionistic fuzzy matrix, intuitionistic fuzzy graph, convergence of intuitionistic fuzzy matrix, convex combination.}
\end {quote}

\section{Introduction}
Intuitionistic fuzzy relations on finite universes can be represented by intuitionistic fuzzy matrix (IFM). The powers of an
IFM play a crucial role in finding the transitive closure of the underlying intuitionistic fuzzy relation. For an IFM $A$, we
mean $A=[\langle a_{ij\mu},a_{ij\nu}\rangle]$, where $a_{ij\mu}$ and $a_{ij\nu}$ are the membership and non-membership values such
that, $0\leq a_{ij\mu}+a_{ij\nu}\leq 1$. Let $A$ be an IFM of order $n$. Given $\lambda \in[0,1]$ and for a non-zero real number
$p$, the maxgeneralized mean-mingeneralized mean operation, denoted by $``\circ"$ for an IFM $A$ can be defined as
\begin{eqnarray}
[A\circ A]_{ij}&=&\left\langle\max\limits_{1\leq t\leq n}\left\{\left(\lambda a_{it\mu}^p+(1-\lambda)a_{tj\mu}^p\right)^{\frac{1}{p}}\right\},
\min\limits_{1\leq t\leq n}\left\{\left(\lambda a_{it\nu}^p+(1-\lambda)a_{tj\nu}^p\right)^{\frac{1}{p}}\right\}\right\rangle,\nonumber\\
&&\forall 1\leq i,j\leq n.
\end{eqnarray}
For $\lambda=1$ and $p=1$, this operator approaches to max-min operator as,
\begin{eqnarray*}
[A\circ A]_{ij}&=&\left\langle\max\limits_{1\leq t\leq n}a_{it\mu},\min\limits_{1\leq t\leq n}a_{it\nu}\right\rangle,\forall 1\leq i,j\leq n.
\end{eqnarray*}
For $\lambda=\frac{1}{2}$ and $p=1$, this operator approaches to maxarithmetic mean-minarithmetic mean operator as,
\begin{eqnarray*}
[A\circ A]_{ij}&=&\left\langle\max\limits_{1\leq t\leq n}\left\{\frac{a_{it\mu}+a_{tj\mu}}{2}\right\},
\min\limits_{1\leq t\leq n}\left\{\frac{a_{it\nu}+a_{tj\nu}}{2}\right\}\right\rangle,\forall 1\leq i,j\leq n.
\end{eqnarray*}
For $\lambda=\frac{1}{2}$ and $p>0$, this operator approaches to max rootpower mean-min rootpower mean operator as,
\begin{eqnarray*}
[A\circ A]_{ij}&=&\left\langle\max\limits_{1\leq t\leq n}\left\{\left(\frac{a_{it\mu}^p+a_{tj\mu}^p}{2}\right)^{\frac{1}{p}}\right\},
\min\limits_{1\leq t\leq n}\left\{\left(\frac{a_{it\nu}^p+a_{tj\nu}^p}{2}\right)^{\frac{1}{p}}\right\}\right\rangle,\forall 1\leq i,j\leq n.
\end{eqnarray*}
For $\lambda\in [0,1]$ and $p=1$, this operator approaches to maxconvex mean-minconvex mean operator as,
\begin{eqnarray*}
[A\circ A]_{ij}&=&\left\langle\max\limits_{1\leq t\leq n}\left\{\lambda a_{it\mu}+(1-\lambda)a_{tj\mu}\right\},
\min\limits_{1\leq t\leq n}\left\{\lambda a_{it\nu}+(1-\lambda)a_{tj\nu}\right\}\right\rangle,\forall 1\leq i,j\leq n.
\end{eqnarray*}
For $\lambda=\frac{1}{2}$ and $p=-1$, this operator approaches to maxharmonic mean-minharmonic mean operator as,
\begin{eqnarray*}
[A\circ A]_{ij}&=&\left\langle\max\limits_{1\leq t\leq n}\left\{\frac{2}{\frac{1}{a_{it\mu}}+\frac{1}{a_{tj\mu}}}\right\},
\min\limits_{1\leq t\leq n}\left\{\frac{2}{\frac{1}{a_{it\nu}}+\frac{1}{a_{tj\nu}}}\right\}\right\rangle,\forall 1\leq i,j\leq n.
\end{eqnarray*}

Thomason's work \cite{tho77} published in 1977 was the first to express the behavior of powers of a fuzzy matrix. The authors showed that
only two consequences exist for the max-min power of a fuzzy matrix \cite{buc01,fan98,fan97}, either converge to an idempotent
matrix or to oscillate with a finite period. Moreover, he established some sufficient conditions to have convergence. Main way
to prove those sufficient conditions was to assume compactness for fuzzy matrix. On the other hand, Hasimoto \cite{has83}
assumed the fuzzy matrix to be transitive to have convergence. As pointed out explicitly by him, either compactness or
transitivity of the fuzzy matrix shall induce convergence because of the monotonicity of its powers. Bourke and Fisher \cite{bou96}
studied the stability analysis of relational matrices combined with the max-min composition and presents an analysis of the
stability of relational matrices combined with the max-product composition. This analysis includes results defining the
convergence properties of the relational matrix and determination of the eigen fuzzy set of the stable matrices.
If different operations are adopted, then the behavior of the limit matrix of the power sequence of fuzzy matrix may be
significantly different \cite{xin92}. For instance, the max-product powers \cite{fan99} of a fuzzy matrix relate to the
notion of asymptotic period \cite{guu01} and the limiting behavior of the consecutive powers can be completely decided
by a Boolean matrix \cite{pan01}. The power convergence of the Boolean matrices was studied by Gregory et al. \cite{gre93}.
They shown that, a binary matrix $A$ is idempotent if and only if it is limit dominating and the number of non-zero
diagonal blocks in its Frobenious normal form equals its column rank. They also give the natural generalization to
matrices with entries from an arbitrary finite Boolean algebra. First time Pal and Khan \cite{pal02} define intuitionistic
fuzzy matrices (IFM). Then Bhowmik and Pal \cite{bho11} first time discuss the convergency of the max-min powers of
an IFM. Latter Pradhan and Pal \cite{pra13} studied maxarithmetic mean-minarithmetic mean power convergence of IFMs.

The works done by Lur et al. \cite{lur07} motivate us to study the power convergence of IFMs under the operation
maxgeneralized mean-mingeneralized mean. Here we consider the weight of the $m$-path of an intuitionistic fuzzy
graph corresponding to the underlying IFM $A$ in such a way that it has both the membership as well as non-membership
values. We also shown that for power convergence of an IFM, only connected intuitionistic fuzzy graph is sufficient.
In this paper, we also test the convergency of the powers of an IFM with the operation, the convex combination of max-min
and maxarithmetic mean-minarithmetic mean. Here we observe that all the results in \cite{lur07} holds for IFMs also.

This paper is organized as follows. In Section 2, definitions of some basic terms are given. In Section 3,
maxgeneralized mean-mingeneralized mean powers of IFMs are defined. It is shown that this power is always convergent and the sequence
$\{A^n_p\}$ converge faster as the value of $p$ increases. Here it is also shown that, the limit of this sequence has the feature
that all elements of each column are identical. In Section 4, the convergence of powers for an IFM with
convex combination of max-min and maxarithmetic mean-minarithmetic mean operations is considered.
Section 5 is for conclusion.

\section{Preliminaries} In this Section, some elementary aspects that are necessary for this paper are introduced.

In fuzzy matrix, the elements of a matrix are the membership degrees only, but in an intuitionistic fuzzy
matrix the membership degree and non-membership degree both are represented, which is defined as follows.
\begin{definition}{\bf(Intuitionistic fuzzy matrices)}\\
An intuitionistic fuzzy matrix $A$ of order $m\times n$ is defined as
$ A=(\langle a_{ij\mu},a_{ij\nu}\rangle)_{m\times n}$ where $a_{ij\mu}$, $a_{ij\nu}$
are called membership and non-membership values of $ij$-th element of $A$, which maintains the
condition $ 0\leq a_{ij\mu}+a_{ij\nu}\leq1$. For simplicity, we write $ A=[a_{ij}]_{m\times n}$,
where $a_{ij}=\langle a_{ij\mu},a_{ij\nu}\rangle$. All elements of an IFM are the members
of $\langle F\rangle=\{\langle a,b\rangle : 0\leq a+b \leq 1\}$.
\end{definition}

One special type of IFM is universal IFM, Which is defined as,
\begin{definition}{\bf(Universal IFM)}\\
An IFM is said to be universal IFM, if all the elements of this matrix are $\langle 1,0\rangle$ and is denoted by $U$.
\end{definition}

Comparison between intuitionistic fuzzy matrices plays an important role in our work, which is defined below.
\begin{definition}{\bf(Dominance of IFM)}\\
Let $A, B\in F_{m\times n}$ such that $A=(\langle a_{ij\mu},a_{ij\nu}\rangle)$ and $B=(\langle b_{ij\mu},b_{ij\nu}\rangle)$,
then we write $A\leq B$ if, $a_{ij\mu}\leq b_{ij\mu}$ and $a_{ij\nu}\geq b_{ij\nu}$ for all $i, j$, and we say that
$A$ is dominated by $B$ or $B$ dominates $A$. $A$ and $B$ are said to be comparable, if either $A\leq B$ or $B\leq A$.
\end{definition}

To compute the $m$-th power of an IFM we consider the weight of a path of length $m$ of an intuitionistic fuzzy graph
$G=(\mu, \nu, V, E)$, where $V$ is the vertex set, $E$ is the edge set, $\mu$ and $\nu$ represent the membership and
the non-membership values of both the vertices and edges respectively. This graph is defined below.

\begin{figure}[h]
\begin{center}
\unitlength 1mm 
\linethickness{0.4pt}
\ifx\plotpoint\undefined\newsavebox{\plotpoint}\fi 
\begin{picture}(54.25,35.5)(0,0)
\put(16.25,28.25){\circle{9.605}}
\put(47.5,28.25){\circle{10.966}}
\put(13.75,7.5){\circle{10.259}}
\put(47.5,7.75){\circle{10.404}}
\put(20.75,29.5){\line(1,0){21.25}}
\put(18.25,7.75){\line(1,0){24.5}}
\put(16,28.25){\makebox(0,0)[cc]{$v_1$}}
\put(47.75,28.75){\makebox(0,0)[cc]{$v_2$}}
\put(53.5,35.5){\makebox(0,0)[cc]{$\langle 0.5,0.5 \rangle$}}
\put(13.75,7.75){\makebox(0,0)[cc]{$v_3$}}
\put(47.5,7.75){\makebox(0,0)[cc]{$v_4$}}
\put(5.5,33.25){\makebox(0,0)[cc]{$\langle 0.6,0.3 \rangle$}}
\put(11.25,.5){\makebox(0,0)[cc]{$\langle 0.7,0.1 \rangle$}}
\put(7.5,18.75){\makebox(0,0)[cc]{$\langle 0.7,0.2 \rangle$}}
\put(30,5.25){\makebox(0,0)[cc]{$\langle 0.6,0.1 \rangle$}}
\put(46.75,.25){\makebox(0,0)[cc]{$\langle 0.5,0.3 \rangle$}}
\put(15.25,24){\line(0,-1){12}}
\put(46.75,23.25){\line(0,-1){10.5}}
\put(30.75,32.25){\makebox(0,0)[cc]{$\langle 0.6,0.3 \rangle$}}
\put(54.25,18.5){\makebox(0,0)[cc]{$\langle 0.5,0.4 \rangle$}}
\end{picture}
\end{center}
\caption{Intuitionistic fuzzy graph}\label{fig1}
\end{figure}
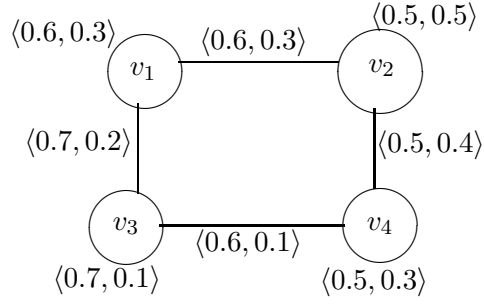

\begin{definition}{\bf(Intuitionistic fuzzy graph)}\\
A graph $G=(\mu, \nu, V, E)$ is said to be max-min intuitionistic fuzzy graph (IFG) if \\
(i) $V=\{v_{1},v_{1},\ldots, v_{n}\}$ such that, $\mu_{1}:V\rightarrow [0,1]$ and $\nu_{1}:V\rightarrow [0,1]$,
denote the degree of membership and the degree of non-membership values of the vertex $v_{i}\in V$ respectively
and $0\leq \mu_{1}(v_{i})+\nu_{1}(v_{i})\leq 1$, for every $v_{i}\in V$, and\\
(ii) $E\subseteq V\times V$ where $\mu_{2}:V\times V\rightarrow [0,1]$ and $\nu_{2}:V\times V\rightarrow [0,1]$ are such that
$\mu_{2}(v_{i},v_{j})\leq max\{\mu_{1}(v_{i}),\mu_{1}(v_{j})\}$ and $\nu_{2}(v_{i},v_{j})\geq min\{\nu_{1}(v_{i}),\nu_{1}(v_{j})\}$,
denotes the membership and non-membership values of the edge $(v_{i},v_{j})\in E$ respectively, where, $0\leq \mu_{2}(v_{i},v_{j})+\nu_{2}(v_{i},v_{j})\leq 1$, for every $(v_{i},v_{j})\in E$.

An IFG $G=(\mu,\nu,V,E)$ is said to be complete if $\mu_{2}(v_{i},v_{j})=max\{\mu_{1}(v_{i}),\mu_{1}(v_{j})\}$ and $\nu_{2}(v_{i},v_{j})= min\{\nu_{1}(v_{i}),\nu_{1}(v_{j})\}$ for all $v_{i},v_{j}\in V$.
\end{definition}
An example of an intuitionistic fuzzy graph with four vertices is shown in Figure \ref{fig1}.

\begin{definition}{\bf (Length of a path)}
A path $P$ in an IFG $G=(\mu, \nu, V, E)$ is said to be of length $m$ or a $m$-path if it is a sequence of $(m+1)$ distinct vertices $v_{0},v_{1},\ldots, v_{m}$ of the vertex set $V$.
\end{definition}
In this paper, weight of the path has an crucial role to find the power of IFM, which is defined as follows.
\begin{definition}{\bf (Wieght of a path)}
Let $G$ be an IFG corresponding to the IFM $A$ and $P_m$ be a path of length $m$ of the edges $(i_{0},i_{1}),(i_{1},i_{2}),\ldots, (i_{m-1},i_{m})$.
Then the weight of $P_m$ is an intuitionistic fuzzy number (IFN) $w(P_m)$, whose membership and non-membership parts are the generalized mean of the
membership and non-membership values of the edges in the said path respectively.
\end{definition}

\section{Maxgeneralized mean-mingeneralized mean powers}
The maxgeneralized mean-mingeneralized mean operation $``\circ"$ between two IFMs $A=[a_{ij}]_{n\times m}$ and $B=[b_{ij}]_{m\times n}$
can be stated as,
\begin{eqnarray}
[A\circ B]_{ij}&=&\left\langle\max\limits_{1\leq t\leq n}\left\{\left(\lambda a_{it\mu}^p+(1-\lambda)b_{tj\mu}^p\right)^{\frac{1}{p}}\right\},
\min\limits_{1\leq t\leq n}\left\{\left(\lambda a_{it\nu}^p+(1-\lambda)b_{tj\nu}^p\right)^{\frac{1}{p}}\right\}\right\rangle,\nonumber\\
&&\forall 1\leq i,j\leq n.
\end{eqnarray}

It follows that $A\circ(B\circ C)$ is not necessarily equal to $(A\circ B)\circ C$, that is, $``\circ"$ is non-associative.
Therefore, the powers $A^{k}$ of $A$ can be defined as $A^{k}=(A^{k-1})\circ A, k=2, 3, \ldots$. Again, $(A^{k-1})\circ A$
may not be equal to $A\circ(A^{k-1})$.

The directed intuitionistic fuzzy graph corresponding to the IFM $A$ of order $n$, is defined by $G=(\mu,\nu,V,E)$ with the vertex
set $V=\{1,2,\ldots, n\}$ and the edge set $E=\{(i,j)\in V\times V|1\leq i,j\leq n\}$. A path of length $k$ is a sequence of
$k$ edges $(i_{0},i_{1}),(i_{1},i_{2}),\ldots, (i_{k-1},i_{k})$ and it is called $k$-path. This is denoted by
$P(i_{0},i_{1},\ldots, i_{k})$. The weight of the path $P(i_{0},i_{1},\ldots, i_{k})$ is denoted by $w(P(i_{0},i_{1},\ldots, i_{k}))$
or simply by $w(P)$, is defined by
\begin{eqnarray*}
 w(P(i_{0},i_{1},\ldots, i_{k}))&=&\left\langle w_{\mu}(P),w_{\nu}(P)\right\rangle, \hspace{0.1cm} \mbox{where}\\
w_{\mu}(P)&=&\Big(\lambda^{k-1}a_{i_0i_1\mu}^p+\lambda^{k-2}(1-\lambda)a_{i_1i_2\mu}^p+\ldots +(1-\lambda)a_{i_{k-1}i_k\mu}^p\Big)^{\frac{1}{p}} \hspace{0.1cm} \mbox{and}\\
w_{\nu}(P)&=&\Big(\lambda^{k-1}a_{i_0i_1\nu}^p+\lambda^{k-2}(1-\lambda)a_{i_1i_2\nu}^p+\ldots +(1-\lambda)a_{i_{k-1}i_k\nu}^p\Big)^{\frac{1}{p}}.
\end{eqnarray*}

A path $P(i_{0},i_{1},\ldots, i_{k})$ is called a critical path from the vertex $i_{0}$ to the vertex $i_{k}$ if
$w(P)=\langle 1,0\rangle$, that is, $\langle a_{i_0i_1\mu},a_{i_0i_1\nu}\rangle=\langle a_{i_1i_2\mu},a_{i_1i_2\nu}\rangle=\dots =\langle a_{i_{k-1}i_k\mu},a_{i_{k-1}i_k\nu}\rangle=\langle 1,0\rangle$. A circuit $C$ of length $k$ is a path $P(i_{0},i_{1},\ldots, i_{k})$ with
$i_{k}=i_{0}$, where $i_{1},i_{2},\ldots, i_{k-1}$ are distinct. A circuit $C$ with $w(C)=\langle 1,0\rangle$ is called a critical circuit and
vertices on critical circuit are called critical vertices.

\begin{theorem}\label{t0} Let $A$ be a square IFM of order $n$. Then the $ij$-th element of the $m$-th power of $A$ will be $[A^{m}]_{ij}=\langle max\{w_{\mu}(P_{m})\},min\{w_{\nu}(P_{m})\}\rangle$, where $P_{m}$ is the $m$-path from the vertex $i$ to the vertex $j$ of the corresponding IFG.
\end{theorem}
{\bf Proof:} Let $W=\langle max\{w_{\mu}(P_{k})\},min\{w_{\nu}(P_{k})\}\rangle$, where $P_{k}$ is the $k$-path from the vertex $i$ to the vertex $j$.
We prove the theorem by mathematical induction on $m$.

The theorem is true for $m=1$. Let us consider that the theorem is true for $m=k-1$ also. Then choose a vertex $s$, $1\leq s\leq n$ such that,
\begin{eqnarray*}
[A^{k-1}]_{is}\circ[A]_{sj}&=&\Big\langle \max \limits_{1\leq t\leq n}\Big(\lambda[A^{k-1}]_{it\mu}^{p}+(1-\lambda)[A]_{tj\mu}^{p}\Big)^{\frac{1}{p}},
\min \limits_{1\leq t\leq n}\Big(\lambda[A^{k-1}]_{it\nu}^{p}+(1-\lambda)[A]_{tj\nu}^{p}\Big)^{\frac{1}{p}}\Big\rangle\\
&=&[A^{k}]_{ij}.
\end{eqnarray*}

By induction hypothesis, there are some $(k-1)$-path $P_{k-1}=P(i_{0}=i,i_{1},\ldots, i_{k-1}=s)$ such that,
\begin{eqnarray*}
[A^{k-1}]_{is}&=&w(P_{k-1})\\
&=&\Big\langle\Big(\lambda^{k-2}a_{ii_1\mu}^{p}+\lambda^{k-3}(1-\lambda)a_{i_1i_2\mu}^{p}+\ldots+(1-\lambda)a_{i_{k-2}i_{k-1}\mu}^{p}\Big)^{\frac{1}{p}},\\
&&\Big(\lambda^{k-2}a_{ii_1\nu}^{p}+\lambda^{k-3}(1-\lambda)a_{i_1i_2\nu}^{p}+\ldots+(1-\lambda)a_{i_{k-2}i_{k-1}\nu}^{p}\Big)^{\frac{1}{p}}\Big\rangle.
\end{eqnarray*}

Let $P_{k}=(i_{0}=i,i_{1},\ldots, i_{k-1}=s,i_{k}=j)$. Then $P_k$ is a $k$-path from the vertex $i$ to the vertex $j$ with
\begin{eqnarray*}
w(P_{k})&=&w(P_{k-1})\circ a_{sj}\\
&=&\Big\langle\Big(\lambda^{k-1}a_{ii_1\mu}^{p}+\lambda^{k-2}(1-\lambda)a_{i_1i_2\mu}^{p}+\ldots+(1-\lambda)a_{sj\mu}^{p}\Big)^{\frac{1}{p}},\\
&&\Big(\lambda^{k-1}a_{ii_1\nu}^{p}+\lambda^{k-2}(1-\lambda)a_{i_1i_2\nu}^{p}+\ldots+(1-\lambda)a_{sj\nu}^{p}\Big)^{\frac{1}{p}}\Big\rangle\\
&=&[A^{k}]_{sj}.
\end{eqnarray*}

This implies,
\begin{eqnarray}
\label{e1}
[A^{k}]_{ij}\leq W.
\end{eqnarray}

On the other hand, let $P_{k}=(i_{0}=i,i_{1},\ldots, i_{k}=j)$ be given arbitrary path. Then putting $P_{k-1}=P(i_{0}=i,i_{1},\ldots, i_{k-1})$ we get,
\begin{eqnarray*}
w(P_{k})&=&\Big\langle\Big(\lambda^{k-1}a_{ii_1\mu}^{p}+\lambda^{k-2}(1-\lambda)a_{i_1i_2\mu}^{p}+\ldots+(1-\lambda)a_{i_{k-1}i_{k}\mu}^{p}\Big)^{\frac{1}{p}},\\
&&\Big(\lambda^{k-1}a_{ii_1\nu}^{p}+\lambda^{k-2}(1-\lambda)a_{i_1i_2\nu}^{p}+\ldots+(1-\lambda)a_{i_{k-1}i_{k}\nu}^{p}\Big)^{\frac{1}{p}}\Big\rangle.
\end{eqnarray*}

By induction hypothesis,
\begin{eqnarray*}
[A^{k-1}]_{i_0i_{k-1}}&\geq& w(P_{k-1}).
\end{eqnarray*}

Hence,
\begin{eqnarray*}
w(P_{k})&\leq& [A^{k-1}]_{i_0i_{k-1}}\circ a_{i_{k-1}i_{k}}\\
&\leq& [A^{k}]_{ij}.
\end{eqnarray*}

This shows that,
\begin{eqnarray}
\label{e2}
[A^{k}]_{ij}\geq W.
\end{eqnarray}

By (\ref{e1}) and (\ref{e2}), the only possibility is, $[A^k]_{ij}=W$.

Hence the assertion is true for $m=k$ also. That is, the assertion is true for any integer $m$.

\begin{theorem} \label{t1}Let $A$ be an IFM of order $n$. Then the maxgeneralized mean-mingeneralized mean powers of $A$ are convergent.
That is, $\lim \limits_{m\rightarrow \infty}A^{m}$ exists and let it be ${\AA}$.\\
Also, for each $1\leq j\leq n$, ${\AA}_{rj}={\AA}_{sj}$, for all $1\leq r,s\leq n$.
\end{theorem}
{\bf Proof:} (\underline{First part}) Let $1\leq r,s\leq n$ be fixed and $P_{m}=(i_{0}=r,i_{1},\ldots, i_{m-1},i_{m}=s)$ be given. We remove the vertex $i_1$
from the path $P_m$ to form the path $P_{m-1}=(i_{0}=r,i_{2},\ldots, i_{m-1},i_{m}=s)$. Then $P_m$ is a $m-$path from the vertex $r$ to the
vertex $s$ and $P_{m-1}$ is a $(m-1)$-path from the vertex $r$ to the vertex $s$. Then,
\begin{eqnarray*}
w(P_m)&=&\langle w_{\mu}(P_m),w_{\nu}(P_m)\rangle\\
&=&\Big\langle\Big(\lambda^{m-1}a_{i_0i_1\mu}^{p}+\lambda^{m-2}(1-\lambda)a_{i_1i_2\mu}^{p}+\ldots+(1-\lambda)a_{i_{m-1}i_{m}\mu}^{p}\Big)^{\frac{1}{p}},\\
&&\Big(\lambda^{m-1}a_{i_0i_1\nu}^{p}+\lambda^{m-2}(1-\lambda)a_{i_1i_2\nu}^{p}+\ldots+(1-\lambda)a_{i_{m-1}i_{m}\nu}^{p}\Big)^{\frac{1}{p}}\Big\rangle  \hspace{0.1cm}\mbox{and}\\
w(P_{m-1})&=&\langle w_{\mu}(P_{m-1}),w_{\nu}(P_{m-1})\rangle\\
&=&\Big\langle\Big(\lambda^{m-2}a_{i_0i_2\mu}^{p}+\lambda^{m-3}(1-\lambda)a_{i_2i_3\mu}^{p}+\ldots+(1-\lambda)a_{i_{m-1}i_{m}\mu}^{p}\Big)^{\frac{1}{p}},\\
&&\Big(\lambda^{m-2}a_{i_0i_2\nu}^{p}+\lambda^{m-3}(1-\lambda)a_{i_2i_3\nu}^{p}+\ldots+(1-\lambda)a_{i_{m-1}i_{m}\nu}^{p}\Big)^{\frac{1}{p}}\Big\rangle.
\end{eqnarray*}

From the above two equalities we can obtain,
\begin{eqnarray*}
w(P_{m})&=&\Big\langle\Big(\lambda^{m-1}a_{i_0i_1\mu}^{p}+\lambda^{m-2}(1-\lambda)a_{i_1i_2\mu}^{p}+\ldots+(1-\lambda)a_{i_{m-1}i_{m}\mu}^{p}\Big)^{\frac{1}{p}},\\
&&\Big(\lambda^{m-1}a_{i_0i_1\nu}^{p}+\lambda^{m-2}(1-\lambda)a_{i_1i_2\nu}^{p}+\ldots+(1-\lambda)a_{i_{m-1}i_{m}\nu}^{p}\Big)^{\frac{1}{p}}\Big\rangle\\
&=&\Big\langle\Big(\lambda^{m-2}a_{i_0i_2\mu}^{p}+\ldots+(1-\lambda)a_{i_{m-1}i_{m}\mu}^{p}
+\lambda^{m-1}a_{i_0i_1\mu}^p+\lambda^{m-2}(1-\lambda)a_{i_1i_2\mu}^{p}-\lambda^{m-2}a_{i_0i_2\mu}^{p}\Big)^{\frac{1}{p}},\\
&&\Big(\lambda^{m-2}a_{i_0i_2\nu}^{p}+\ldots+(1-\lambda)a_{i_{m-1}i_{m}\nu}^{p}
+\lambda^{m-1}a_{i_0i_1\nu}^p+\lambda^{m-2}(1-\lambda)a_{i_1i_2\nu}^{p}-\lambda^{m-2}a_{i_0i_2\nu}^{p}\Big)^{\frac{1}{p}}\Big\rangle\\
&\leq&w(P_{m-1})+\lambda^{\frac{m-2}{p}}\langle 1,0\rangle \hspace{0.1cm}[\mbox{ as $\max\limits_{1\leq i,j\leq n}a_{ij\mu}\leq 1$ and $\min\limits_{1\leq i,j\leq n}a_{ij\nu}\geq 0$}].
\end{eqnarray*}

This implies, with the help of Theorem \ref{t0}
\begin{eqnarray}\label{e3}
[A^{m}]_{rs}\leq [A^{m-1}]_{rs}+\lambda^{\frac{m-2}{p}}\langle 1,0\rangle.
\end{eqnarray}

On the other hand, let $P_{m-1}=(i_{0}=r,i_{2},\ldots, i_{m-1},i_{m}=s)$ be a $(m-1)$-path from the vertex $r$ to the vertex $s$.
Choose $1\leq i_{1}\leq n$. Let $P_{m}=(i_{0}=r,i_{1},\ldots, i_{m-1},i_{m}=s)$, then $P_m$ is a $m-$path from the vertex $r$
to the vertex $s$. Observe that,
\begin{eqnarray*}
w(P_{m-1})&=&\Big\langle\Big(\lambda^{m-2}a_{i_0i_2\mu}^{p}+\lambda^{m-3}(1-\lambda)a_{i_2i_3\mu}^{p}+\ldots+(1-\lambda)a_{i_{m-1}i_{m}\mu}^{p}\Big)^{\frac{1}{p}},\\
&&\Big(\lambda^{m-2}a_{i_0i_2\nu}^{p}+\lambda^{m-3}(1-\lambda)a_{i_2i_3\nu}^{p}+\ldots+(1-\lambda)a_{i_{m-1}i_{m}\nu}^{p}\Big)^{\frac{1}{p}}\Big\rangle\\
&=&\Big\langle\Big(\lambda^{m-1}a_{i_0i_1\mu}^{p}+\ldots+(1-\lambda)a_{i_{m-1}i_{m}\mu}^{p}
+\lambda^{m-2}a_{i_0i_2\mu}^p-\lambda^{m-1}a_{i_0i_1\mu}^{p}-\lambda^{m-2}(1-\lambda)a_{i_1i_2\mu}^{p}\Big)^{\frac{1}{p}},\\
&&\Big(\lambda^{m-1}a_{i_0i_1\nu}^{p}+\ldots+(1-\lambda)a_{i_{m-1}i_{m}\nu}^{p}
+\lambda^{m-2}a_{i_0i_2\nu}^p+\lambda^{m-1}a_{i_0i_1\nu}^{p}-\lambda^{m-2}(1-\lambda)a_{i_1i_2\nu}^{p}\Big)^{\frac{1}{p}}\Big\rangle\\
&\leq&w(P_{m})+\lambda^{\frac{m-2}{p}}\langle 1,0\rangle \hspace{0.1cm}[\mbox{ as $\max\limits_{1\leq i,j\leq n}a_{ij\mu}\leq 1$ and $\min\limits_{1\leq i,j\leq n}a_{ij\nu}\geq 0$}].
\end{eqnarray*}

This implies,
\begin{eqnarray}\label{e4}
 [A^{m-1}]_{rs}\leq [A^{m}]_{rs}+\lambda^{\frac{m-2}{p}}\langle 1,0\rangle.
\end{eqnarray}

From (\ref{e3}) and (\ref{e4}), we obtain $|[A^m]_{rs}-[A^{m-1}]_{rs}|\leq \lambda^{\frac{m-2}{p}}\langle 1,0\rangle$.

Let $N$ be a fixed natural number and for all $m\geq N$,
\begin{eqnarray*}
|[A^m]_{rs}-[A^N]_{rs}|&\leq& |[A^m]_{rs}-[A^{m-1}]_{rs}|+|[A^{m-1}]_{rs}-[A^{m-2}]_{rs}|+\ldots +|[A^{N+1}]_{rs}-[A^{N}]_{rs}| \\
&\leq& \lambda^{\frac{m-2}{p}}\langle 1,0\rangle+\lambda^{\frac{m-3}{p}}\langle 1,0\rangle+\ldots+\lambda^{\frac{N-1}{p}}\langle 1,0\rangle\\
&\leq& \lambda^{\frac{N-1}{p}}\Big(\frac{\langle 1,0\rangle}{1-\lambda}\Big).
\end{eqnarray*}

Since $0\leq \lambda\leq1$, we have the sequence $\{[A^{m}]_{rs}\}$ is a Cauchy sequence and hence convergent. That imply, $\lim \limits_{m\rightarrow \infty}A^{m}={\AA}$.

(\underline{Second part}) Let $P_{m}(rj)=(i_{0}=r,i_{1},\ldots, i_{m-1},i_{m}=j)$ be a $m$-path from the vertex $r$ to the vertex $j$ and
$P_{m}(sj)=(i_{0}=s,i_{1},\ldots, i_{m-1},i_{m}=j)$ be another $m$-path from the vertex $s$ to the vertex $j$. Then,
\begin{eqnarray*}
w(P_m(rj))&=&\Big\langle\Big(\lambda^{m-1}a_{ri_1\mu}^{p}+\lambda^{m-2}(1-\lambda)a_{i_1i_2\mu}^{p}+\ldots+(1-\lambda)a_{i_{m-1}i_{m}\mu}^{p}\Big)^{\frac{1}{p}},\\
&&\Big(\lambda^{m-1}a_{ri_1\nu}^{p}+\lambda^{m-2}(1-\lambda)a_{i_1i_2\nu}^{p}+\ldots+(1-\lambda)a_{i_{m-1}i_{m}\nu}^{p}\Big)^{\frac{1}{p}}\Big\rangle  \hspace{0.1cm}\mbox{and}\\
w(P_m(sj))&=&\Big\langle\Big(\lambda^{m-1}a_{si_1\mu}^{p}+\lambda^{m-2}(1-\lambda)a_{i_1i_2\mu}^{p}+\ldots+(1-\lambda)a_{i_{m-1}i_{m}\mu}^{p}\Big)^{\frac{1}{p}},\\
&&\Big(\lambda^{m-1}a_{si_1\nu}^{p}+\lambda^{m-2}(1-\lambda)a_{i_1i_2\nu}^{p}+\ldots+(1-\lambda)a_{i_{m-1}i_{m}\nu}^{p}\Big)^{\frac{1}{p}}\Big\rangle.
\end{eqnarray*}
Now,
\begin{eqnarray*}
w(P_m(rj))&=&\Big\langle\Big(\lambda^{m-1}a_{ri_1\mu}^{p}+\ldots+(1-\lambda)a_{i_{m-1}i_{m}\mu}^{p}
+\lambda^{m-1}a_{ri_1\mu}^p-\lambda^{m-1}a_{si_1\mu}^p\Big)^{\frac{1}{p}},\\
&&\Big(\lambda^{m-1}a_{ri_1\nu}^{p}+\ldots+(1-\lambda)a_{i_{m-1}i_{m}\nu}^{p}+\lambda^{m-1}a_{ri_1\nu}^p-\lambda^{m-1}a_{si_1\nu}^p\Big)^{\frac{1}{p}}\Big\rangle\\
&\leq&w(P_m(sj))+\lambda^{\frac{m-1}{p}}\langle 1,0\rangle \hspace{0.1cm}[\mbox{ as $\max\limits_{1\leq i,j\leq n}a_{ij\mu}\leq 1$ and $\min\limits_{1\leq i,j\leq n}a_{ij\nu}\geq 0$}]\\
\mbox{or}, [A^{m}]_{rj}&\leq&[A^m]_{rj}+\lambda^{\frac{m-1}{p}}\langle 1,0\rangle \hspace{0.1cm}[\mbox{by Theorem $\ref{t0}$}].
\end{eqnarray*}
Similarly, $[A^{m}]_{rj}\leq[A^m]_{rj}+\lambda^{\frac{m-1}{p}}\langle 1,0\rangle$.

From the above two inequalities, $|[A^{m}]_{rj}-[A^{m}]_{sj}|\leq \lambda^{\frac{m-1}{p}}\langle 1,0\rangle$.
Now as, $\lim \limits_{m\rightarrow \infty}A^{m}={\AA}$, we can obtain ${\AA}_{rj}={\AA}_{sj}$.

\begin{theorem}\label{t2}
Let $A$ be a square IFM of order $n$ and the powers of it converge to ${\AA}$. Then all entries in the $j$-th column of ${\AA}$ will be $\langle 1,0\rangle$,
if and only if there is a critical path in the IFG $G$ from a critical vertex to the vertex $j$.
\end{theorem}
{\bf Proof:} The condition is necessary.\\
Let $X=\langle x_{ij\mu},x_{ij\nu}\rangle= \Big\langle \max \limits_{1\leq i,j\leq n} a_{ij\mu},\min \limits_{1\leq i,j\leq n}a_{ij\nu}\Big \rangle$, then $x_{ij\mu}\in [0,1]$ and $x_{ij\nu}\in [0,1]$.

Suppose that there is no critical path in $G$ from a critical vertex to the vertex $j$. Let $P_m=(i_{0},i_{1},\ldots, i_{m-1},i_{m}=j)$
be a $m$-path from the vertex $i_0$ to the vertex $j$ with $m\geq n$.

Let us we claim that, the product $\langle a_{i_{m-n}i_{m-n+1}\mu},a_{i_{m-n}i_{m-n+1}\nu}\rangle . \langle a_{i_{m-n+1}i_{m-n+2}\mu},a_{i_{m-n+1}i_{m-n+2w}\nu}\rangle .\\ \ldots  .\langle a_{i_{m-1}i_{m}\mu},a_{i_{m-1}i_{m}\nu}\rangle$ is dominated by $\langle 1,0\rangle$.\\
If $\langle a_{i_{m-n}i_{m-n+1}\mu},a_{i_{m-n}i_{m-n+1}\nu}\rangle . \langle a_{i_{m-n+1}i_{m-n+2}\mu},a_{i_{m-n+1}i_{m-n+2w}\nu}\rangle . \ldots. \langle a_{i_{m-1}i_{m}\mu},a_{i_{m-1}i_{m}\nu}\rangle = \langle 1,0\rangle$ then it imply,
$\langle a_{i_{m-n}i_{m-n+1}\mu},a_{i_{m-n}i_{m-n+1}\nu}\rangle = \langle a_{i_{m-n+1}i_{m-n+2}\mu},a_{i_{m-n+1}i_{m-n+2w}\nu}\rangle = \ldots\\
 = \langle a_{i_{m-1}i_{m}\mu},a_{i_{m-1}i_{m}\nu}\rangle = \langle 1,0\rangle$.

Since $\{i_{m-n},i_{m-n+1},\ldots, i_m\}\subset \{1,2,\ldots, n\}$ with $(n+1)$ elements, there are $m-n\leq r<s<m$ such that $i_r=i_s$. In this situation, the vertex $i_r$ is a critical vertex. If we let $P'=(i_r,i_{r+1},\ldots, i_m=j)$, then $P'$ will be a path from a critical vertex $i_r$ to vertex $j$, a contradiction. Hence we have,\\
 $\langle a_{i_{m-n}i_{m-n+1}\mu},a_{i_{m-n}i_{m-n+1}\nu}\rangle . \langle a_{i_{m-n+1}i_{m-n+2}\mu},a_{i_{m-n+1}i_{m-n+2w}\nu}\rangle . \ldots. \langle a_{i_{m-1}i_{m}\mu},a_{i_{m-1}i_{m}\nu}\rangle$ is dominated by $\langle 1,0\rangle$.

Therefore, there exists at least one $\langle a_{i_{q}i_{q+1}\mu},a_{i_{q}i_{q+1}\nu}\rangle$, that dominated by  $\langle x_{ij\mu},x_{ij\nu}\rangle$ for some $m-n\leq q\leq m-1$. Thus,
\begin{eqnarray*}
w(P(i_0,i_1,\ldots,i_m=j))&=&\Big\langle\Big(\lambda^{m-1}a_{i_0i_1\mu}^{p}+\cdots+\lambda^{m-q-1}(1-\lambda)a_{i_qi_{q+1}\mu}^{p}
+\cdots+(1-\lambda)a_{i_{m-1}i_m\mu}^p\Big)^{\frac{1}{p}},\\
&&\Big(\lambda^{m-1}a_{i_0i_1\nu}^{p}+\cdots+\lambda^{m-q-1}(1-\lambda)a_{i_qi_{q+1}\nu}^{p}
+\cdots+(1-\lambda)a_{i_{m-1}i_m\nu}^p\Big)^{\frac{1}{p}}\Big\rangle\\
&\leq&\Big\{\lambda^{m-1}+\lambda^{m-2}(1-\lambda)+\cdots+\lambda^{m-q}(1-\lambda)+\cdots+(1-\lambda)\Big\}^{\frac{1}{p}}\langle 1,0\rangle\\
&=&\Big\{\frac{(1-\lambda)(1-\lambda^m)}{(1-\lambda)}\Big\}^{\frac{1}{p}}\langle 1,0\rangle\\
&=&(1-\lambda^m)^{\frac{1}{p}}\langle 1,0\rangle.
\end{eqnarray*}

This leads, $[{\AA}]_{i_oj}=\lim \limits_{m\rightarrow \infty}[A^{m}]_{i_0j}\leq (1-\lambda^m)^{\frac{1}{p}}\langle 1,0\rangle$, which contradicts $[{\AA}]_{i_oj}=\langle 1,0\rangle$.

That imply, our assumption is wrong, that is, there is a critical path in $G$ from a critical vertex to the vertex $j$.\\
The condition is sufficient.\\
Let $P^{*}=(i_{0},i_{1},\ldots, i_{s-1},i_{s}=j)$ be a critical path from a critical vertex $i_0$ to the vertex $j$ and let $C=(r_0=i_0,r_1,\ldots, r_h=i_0)$ be a critical circuit of length $h$. For $m$ large enough, let $0\leq k\leq h-1$ such that $m-s=hl+k$ for some positive integer $l$. Choose $1\leq t \leq n$, let $C_1=(j_0=t,j_1,\ldots,j_m=i_0)$ be a $k$-path from the vertex $t$ to the vertex $i_0$.

Then, $P=C_{1}+C+C+\ldots+C+P^{*}$ is a $m$-path from the vertex $t$ to the vertex $j$ and we have,
\begin{eqnarray*}
[A^m]_{tj}&\geq&w(P)\\
&=&\Big\langle\Big[\lambda^{m-1}a_{tj_1\mu}^p+\lambda^{m-2}(1-\lambda)a_{j_1j_2\mu}^p+\ldots+\lambda^{m-k}(1-\lambda)a_{j_{m-1}j_m\mu}^p\\
&&+\lambda^{m-k-1}(1-\lambda)+\ldots+(1-\lambda)\Big]^{\frac{1}{p}},\\
&&\Big[\lambda^{m-1}a_{tj_1\mu}^p+\lambda^{m-2}(1-\lambda)a_{j_1j_2\mu}^p+\ldots+\lambda^{m-k}(1-\lambda)a_{j_{m-1}j_m\mu}^p\\
&&+\lambda^{m-k-1}(1-\lambda)+\ldots+(1-\lambda)\Big]^{\frac{1}{p}}\Big\rangle\\
&=&(1-\lambda^m)^{\frac{1}{p}}\langle 1,0\rangle.
\end{eqnarray*}
As $m$ is fixed and $\lambda\in[0,1]$, we conclude that, $\lim \limits_{m\rightarrow \infty}[A^{m}]_{tj}=\langle 1,0\rangle$.

\begin{theorem}\label{t3}
Let $A$ be an $n\times n$ IFM and ${\AA}=\lim \limits_{m\rightarrow \infty}A^{m}$. Then the limit IFM will be the universal IFM. That is,
${\AA}=U$ if and only if there exists an entry $\langle 1,0\rangle$ in each column of $A$.
\end{theorem}
{\bf Proof:} The above theorem can be proved by the  help of Theorem \ref{t2}. Then, it is sufficient to show that for each $j$ there is a critical path from a critical vertex to the vertex $j$. Since each column of $A$ contains $\langle 1,0\rangle$, for this $j$ there is a vertex $i_{0}$, $1\leq i_0 \leq n$ such that $\langle a_{i_0j\mu},a_{i_0j\nu}\rangle=\langle 1,0\rangle$. For this $i_0$ there is a vertex $i_{1}$, $1\leq i_1 \leq n$ such that $\langle a_{i_1i_0\mu},a_{i_1i_0\nu}\rangle=\langle 1,0\rangle$. Continuing by this way, we obtain $1\leq j,i_0,\ldots,i_{n-1}\leq n$ such that $\langle a_{i_0j\mu},a_{i_0j\nu}\rangle=\langle 1,0\rangle$ and $\langle a_{i_ti_{t-1}\mu},a_{i_ti_{t-1}\nu}\rangle=\langle 1,0\rangle$ for all $t=0,1,\ldots, n-1$. As $1\leq j,i_0,\ldots,i_{n-1}\leq n$, there is $0\leq r\leq n-1$ such that $i_r\in \{i_{r-1},i_{r-2},\ldots, i_0,j\}$. Therefore, the vertex $i_r$ is a critical vertex and $P=(i_r,i_{r-1},\ldots, i_0,j)$ is the required path.

\begin{example}
Let us consider the IFM $A=$$\left[
   \begin{array}{ccc}
   \langle 1,0 \rangle & \langle 0.5,0.4 \rangle & \langle 0,1 \rangle\\
   \langle 0,1 \rangle & \langle 0.6,0.3 \rangle & \langle 1,0 \rangle \\
   \langle 1,0 \rangle & \langle 1,0 \rangle & \langle 0,1 \rangle\\
   \end{array}
\right ]$.

Then the directed IFG corresponding to the IFM $A$ is given in Figure \ref{fig2}.

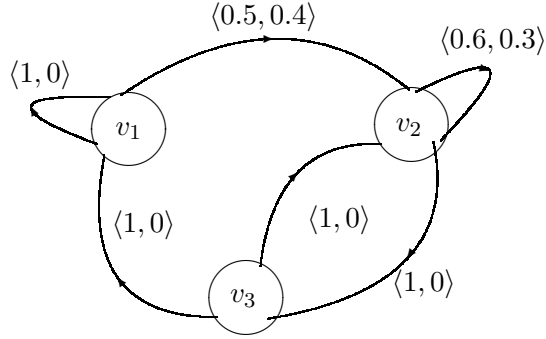
\begin{figure}[h]
\begin{center}
\unitlength 1mm 
\linethickness{0.4pt}
\ifx\plotpoint\undefined\newsavebox{\plotpoint}\fi 
\begin{picture}(68.625,58.375)(0,0)
\put(12.5,39.25){\circle{9.179}}
\put(50.089,39.839){\circle{9.179}}
\put(28.089,16.839){\circle{9.179}}
\put(31.75,51.25){\vector(1,0){.07}}\qbezier(11.5,43.75)(32.875,58.375)(49.75,44.5)
\put(49.5,22){\vector(-3,-4){.07}}\qbezier(53,37.5)(57,18.25)(31,14)
\put(12.5,39){\makebox(0,0)[cc]{$v_1$}}
\put(49.75,39.5){\makebox(0,0)[cc]{$v_2$}}
\put(27.75,16.75){\makebox(0,0)[cc]{$v_3$}}
\put(51.75,18.75){\makebox(0,0)[cc]{$\langle 1,0\rangle$}}
\put(40.5,27.25){\makebox(0,0)[cc]{$\langle 1,0\rangle$}}
\put(11,19.5){\vector(-3,4){.07}}\qbezier(24.25,14.25)(5.25,14)(9.25,35.75)
\put(-.5,42.25){\vector(-1,2){.07}}\qbezier(8.25,37.25)(-10.125,44.125)(10,43.5)
\put(60.5,47){\vector(3,-1){.07}}\qbezier(50.75,44.25)(68.625,53)(54,37.75)
\put(35,33.5){\vector(1,1){.07}}\qbezier(30,21)(32,37.875)(46,37.25)
\put(.75,46.5){\makebox(0,0)[cc]{$\langle 1,0\rangle$}}
\put(30.5,54.25){\makebox(0,0)[cc]{$\langle 0.5,0.4\rangle$}}
\put(61,51){\makebox(0,0)[cc]{$\langle 0.6,0.3\rangle$}}
\put(14.5,26.5){\makebox(0,0)[cc]{$\langle 1,0\rangle$}}
\end{picture}
\end{center}
\caption{Directed IFG G}\label{fig2}
\end{figure}

One of its critical circuit is $v_2\rightarrow v_3\rightarrow v_2$ (see Figure \ref{fig3}) and the vertex $v_1$ has a self-loop. Then the set of all critical vertices in the directed intuitionistic fuzzy graph $G$ is $\{v_1,v_2,v_3\}$.

\begin{figure}[h]
\begin{center}
\unitlength 1mm 
\linethickness{0.4pt}
\ifx\plotpoint\undefined\newsavebox{\plotpoint}\fi 
\begin{picture}(37.947,16.5)(0,0)
\put(2.5,6.25){\circle{8.732}}
\put(33.25,7.5){\circle{9.394}}
\put(17.5,1.25){\vector(1,0){.07}}\qbezier(5.5,3.5)(17.125,-1.25)(30.25,4)
\put(2.25,6.25){\makebox(0,0)[cc]{$v_2$}}
\put(33.5,7.25){\makebox(0,0)[cc]{$v_3$}}
\put(19.25,13){\vector(-1,0){.07}}\qbezier(29.25,9.75)(21.125,16.5)(5.5,9.25)
\end{picture}
\end{center}
\caption{A critical circuit in Example 3.5}\label{fig3}
\end{figure}
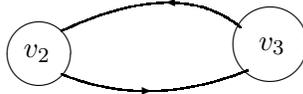

Now, for $\lambda=0.6$ and $p=1$ it is seen that the limit matrix is,\\
$\hat{A}=A^{25}=$$\left[
   \begin{array}{ccc}
   \langle 1,0 \rangle & \langle 1,0 \rangle & \langle 1,0 \rangle\\
   \langle 1,0 \rangle & \langle 1,0 \rangle & \langle 1,0 \rangle \\
   \langle 1,0 \rangle & \langle 1,0 \rangle & \langle 1,0 \rangle\\
   \end{array}
\right ]$.

Here all entries in columns $1$, $2$ and $3$ are $\langle 1,0\rangle$. So $\hat{A}=A^{25}=U$ (the universal IFM).
\end{example}

In general, it is not true that the sequence of powers of an IFM converge faster when we have bigger $p$. Though, we can
provide a sufficient condition for which the power sequence converge faster for bigger $p$. For a fixed power $n$, we
denote $A_p^n$ to emphasize the parameter $p$ being used.

\begin{theorem}\label{t4}
Let $p$ and $q$ be two real numbers with $p\leq q$. For any non-negative integer $n$, we have $A_p^n\leq A_q^n$.
\end{theorem}
{\bf Proof:} For a particular $n$-path $P_n$, we can write
\begin{eqnarray*}
A_p^n&=&\Big\langle\Big(\lambda^{n-1}a_{i_0i_1\mu}^p+\lambda^{n-2}(1-\lambda)a_{i_1i_2\mu}^p+\ldots+(1-\lambda)a_{i_{n-1}i_n\mu}^p\Big)^{\frac{1}{p}},\\
&&\Big(\lambda^{n-1}a_{i_0i_1\nu}^p+\lambda^{n-2}(1-\lambda)a_{i_1i_2\nu}^p+\ldots+(1-\lambda)a_{i_{n-1}i_n\nu}^p\Big)^{\frac{1}{p}}\Big\rangle\\
&\leq&\Big\{\lambda^{n-1}+\lambda^{n-2}(1-\lambda)+\ldots+(1-\lambda)\Big\}^{\frac{1}{p}}\langle 1,0\rangle\\
&=&(1-\lambda^n)^{\frac{1}{p}}\langle 1,0\rangle
\end{eqnarray*}

Similarly, $A_q^n\leq (1-\lambda)^{\frac{1}{q}}\langle 1,0\rangle$.

Now, as $0\leq \lambda<1$ and $p\leq q$, so $(1-\lambda^n)^{\frac{1}{p}}\langle 1,0\rangle\leq (1-\lambda)^{\frac{1}{q}}\langle 1,0\rangle$.

Hence, we can write, $A_p^n\leq A_q^n$.

\begin{theorem}\label{t5}
If each column of the IFM $A$ contains the entry $\langle1,0\rangle$ or for each vertex $j$ there is critical path from a critical vertex
to the vertex $j$, then the powers $A_n^n$ converge to $U$ faster than $A_p^n$, where $p\leq q$.
\end{theorem}
{\bf Proof:} From Theorems \ref{t2} and \ref{t3}, both the conditions imply the powers $A^n$ converge to $U$. Again from Theorem \ref{t4},
since $p\leq q$, we have $A_p^n\leq A_q^n\leq U$. This implies the powers $A_q^n$ converge faster than $A_p^n$.

\section{Convex combination of max-min and maxarithmetic mean-minarithmetic mean operations}
In this Section, we describe the convergence of powers of IFMs with convex combination of max-min and
maxarithmetic mean-minarithmetic mean operations.

Let $A$ be an IFM of order $n$. Given $\lambda\in [0,1]$, the convex combination of max-min and
maxarithmetic mean-minarithmetic mean operations, denoted by ``*", for the IFM $A$ can be defined as,
\begin{eqnarray}\label{e5}
[A*A]_{ij}&=&\Big\langle\max\limits_{1\leq t\leq n}\Big\{\lambda\min(a_{it\mu},a_{tj\mu})+(1-\lambda)\frac{a_{it\mu}+ a_{tj\mu}}{2}\Big\},\nonumber\\
&&\min\limits_{1\leq t\leq n}\Big\{\lambda\max(a_{it\nu},a_{tj\nu})+(1-\lambda)\frac{a_{it\nu}+ a_{tj\nu}}{2}\Big\}\Big\rangle \nonumber\\
&&\forall 1\leq i,j\leq n.
\end{eqnarray}

We observe that if $\lambda=1$ then the operation ``*" becomes the commonly seen max-min operation. On the other hand if $\lambda=0$, then ``*"
becomes the maxarithmetic mean-minarithmetic mean operation.

\begin{definition}{\bf(Scalar multiplication)}\\
Let $\lambda\in [0,1]$ and $a=\langle a_\mu,a_\nu\rangle$ be an intuitionistic fuzzy number (IFN). Then the scalar multiplication of $\lambda$
and $a$ is denoted by $\lambda a$ and can be defined as, $\lambda a=\langle \lambda a_\mu,(1-\lambda)a_\nu\rangle$.
\end{definition}

\begin{definition}{\bf(Difference of two IFNs)}\\
Let $a$ and $b$ be two IFNs, such that, $a$ is dominated by $b$. Then the difference of $a$ from $b$ is denoted by $(b-a)$ and is defined as,
$b-a=\langle b_\mu, b_\nu\rangle - \langle a_\mu, a_\nu\rangle=\langle b_\mu-a_\mu,a_\nu-b_\nu\rangle$.
\end{definition}

Let us consider two IFNs $a$ and $b$, such that, $a\leq b$ and set $\alpha=\frac{1+\lambda}{2}$ $(0\leq \lambda <1)$, then we have $\frac{1}{2}\leq \alpha<1$.
Then from the Equation (\ref{e5}) we write $$a*b=\left\langle \alpha a_\mu+(1-\alpha)b_\mu,\alpha a_\nu+(1-\alpha)b_\nu\right\rangle.$$
We also observe that, for three IFNs $a,b,c$ with $0\leq\lambda<1$ and $a<b$,
\begin{eqnarray}\label{e6}
b*c-a*c\leq \frac{1+\lambda}{2}(b-a).
\end{eqnarray}
\begin{example}
When $c\leq a<b$, let $a=\langle 0.7,0.3\rangle$, $b=\langle 0.8,0.1\rangle$ and $c=\langle 0.6,0.3\rangle$ with $\lambda=0.4$. Then $\alpha=0.7$.
Now
\begin{eqnarray*}
b*c-a*c&=&\langle \alpha c_\mu+(1-\alpha)b_\mu,\alpha c_\nu+(1-\alpha)b_\nu\rangle - \langle \alpha c_\mu+(1-\alpha)a_\mu,\alpha c_\nu+(1-\alpha)a_\nu\rangle\\
&=&\langle 0.66,0.24\rangle - \langle 0.63,0.30\rangle\\
&=&\langle 0.03,0.06\rangle \hspace{0.1cm}\mbox{and}\\
\alpha(b-a)&=&0.7\langle 0.1,0.2\rangle\\
&=&\langle 0.07,0.06\rangle.
\end{eqnarray*}
In this case, $b*c-a*c<\alpha(b-a)$.

When $a\leq c\leq b$, let $a=\langle 0.5,0.4\rangle$, $b=\langle 0.8,0.1\rangle$ and $c=\langle 0.8,0.1\rangle$ with $\lambda=0.4$. Then $\alpha=0.7$.
Now
\begin{eqnarray*}
b*c-a*c&=&\langle \alpha c_\mu+(1-\alpha)b_\mu,\alpha c_\nu+(1-\alpha)b_\nu\rangle - \langle \alpha a_\mu+(1-\alpha)c_\mu,\alpha a_\nu+(1-\alpha)c_\nu\rangle\\
&=&\langle 0.80,0.17\rangle - \langle 0.59,0.37\rangle\\
&=&\langle 0.21,0.20\rangle \hspace{0.1cm}\mbox{and}\\
\alpha(b-a)&=&0.7\langle 0.3,0.3\rangle\\
&=&\langle 0.21,0.09\rangle.
\end{eqnarray*}
In this case, $b*c-a*c<\alpha(b-a)$ also.

When $a<b<c$, let $a=\langle 0.6,0.3\rangle$, $b=\langle 0.7,0.2\rangle$ and $c=\langle 0.8,0.1\rangle$ with $\lambda=0.4$. Then $\alpha=0.7$.
Now
\begin{eqnarray*}
b*c-a*c&=&\langle \alpha b_\mu+(1-\alpha)c_\mu,\alpha b_\nu+(1-\alpha)c_\nu\rangle - \langle \alpha a_\mu+(1-\alpha)c_\mu,\alpha a_\nu+(1-\alpha)c_\nu\rangle\\
&=&\langle 0.07,0.07\rangle \hspace{0.1cm}\mbox{and}\\
\alpha(b-a)&=&0.7\langle 0.1,0.1\rangle\\
&=&\langle 0.07,0.03\rangle.
\end{eqnarray*}
In this case, $b*c-a*c<\alpha(b-a)$.

So for all cases $b*c-a*c\leq\frac{1+\lambda}{2}(b-a)$ for $0\leq\lambda<1$.
\end{example}

Now, the directed intuitionistic fuzzy graph corresponding to the IFM $A$ of order $n$, is defined by $G=(\mu,\nu,V,E)$
with the vertex set $V=\{1,2,\ldots, n\}$ and the edge set $E=\{(i,j)\in V\times V|1\leq i,j\leq n\}$.
Here the weight of the path $P(i_{0},i_{1},\ldots, i_{k})$ is defined by $w(P(i_{0},i_{1},\ldots, i_{k}))=a_{i_0i_1}*a_{i_1i_2}*\ldots*a_{i_{k-1}i_k}$,
where $a_{i_{m-1}i_m}=\langle a_{i_{m-1}i_m\mu},a_{i_{m-1}i_m\nu}\rangle$ is the membership and non-membership values of the edge $(i_{m-1},i_m)$.

\begin{lemma}\label{L2} Let $A$ be a square IFM of order $n$. Then the $m$-th power of $A$ with convex combination of max-min and maxarithmetic mean-minarithmetic mean operation will be  $[A^{m}]_{ij}=\langle max\{w_{\mu}(P_{m})\},min\{w_{\nu}(P_{m})\}\rangle$, where $P_{m}$
is a $m$-path from the vertex $i$ to the vertex $j$ in the IFG $G$.
\end{lemma}
{\bf Proof:} Let $\beta=\langle max\{w_{\mu}(P_{k})\},min\{w_{\nu}(P_{k})\}\rangle$, where $P_{k}$ is a $k$-path from the vertex $i$ to the vertex $j$.
We proceed by induction on $m$.

The assertion is true for $m=1$. Let us consider that the assertion is true for $m=k-1$. Choose $1\leq s\leq n$ such that,
\begin{eqnarray*}
[A^{k-1}]_{is}\circ[A]_{sj}&=&\Big\langle\max\limits_{1\leq t\leq n}\Big\{\lambda\min\Big([A^{k-1}]_{it\mu},[A]_{tj\mu}\Big)+
(1-\lambda)\frac{[A^{k-1}]_{it\mu}+ [A]_{tj\mu}}{2}\Big\},\\
&&\min\limits_{1\leq t\leq n}\Big\{\lambda\max\Big([A^{k-1}]_{it\nu},[A]_{tj\nu}\Big)+
(1-\lambda)\frac{[A^{k-1}]_{it\nu}+ [A]_{tj\nu}}{2}\Big\}\Big\rangle\\
&=&[A^{k}]_{ij}.
\end{eqnarray*}

By induction hypothesis, there are some $(k-1)$-path $P_{k-1}=P(i_{0}=i,i_{1},\ldots, i_{k-1}=s)$ such that,
$[A^{k-1}]_{is}=w(P_{k-1})$. Hence, $[A^k]_{ij}=w(P_{k-1})*a_{sj}=w(P_k)$ where, $P_{k}=(i_{0}=i,i_{1},\ldots, i_{k-1}=s,i_{k}=j)$,
is a $k$-path from the vertex $i$ to the vertex $j$.

This implies,
\begin{eqnarray}
\label{e7}
[A^{k}]_{ij}\leq \beta.
\end{eqnarray}

On the other hand, let $P_{k}=(i_{0}=i,i_{1},\ldots, i_{k}=j)$ be given arbitrary path. Then putting $P_{k-1}=P(i_{0}=i,i_{1},\ldots, i_{k-1})$ we get,
$w(P_{k})=a_{i_0i_1}*a_{i_1i_2}*\ldots*a_{i_{k-1}i_k}$.

By induction hypothesis,
$[A^{k-1}]_{i_0i_{k-1}}\geq a_{i_0i_1}*a_{i_1i_2}*\ldots*a_{i_{k-2}i_{k-1}}$.

Hence, $w(P_{k})\leq [A^{k-1}]_{i_0i_{k-1}}*a_{i_{k-1}i_{k}}\leq [A^k]_{ij}$.

This shows that,
\begin{eqnarray}
\label{e8}
[A^{k}]_{ij}\geq \beta.
\end{eqnarray}

By (\ref{e7}) and (\ref{e8}), the only possibility is, $[A^k]_{ij}=\beta$.

Hence, the assertion is true for $m=k$ also. That is, the assertion is true for any integer $m$.

\begin{theorem} \label{t6}Let $A$ be an IFM of order $n$. Then the powers of $A$ with respect to the operation $``*"$ converges and let the limit matrix
be ${\AA}$. That is, $\lim \limits_{m\rightarrow \infty}A^{m}={\AA}$.\\
Farther more, for each $1\leq j\leq n$, ${\AA}_{rj}={\AA}_{sj}$, for all $1\leq r,s\leq n$.
\end{theorem}
{\bf Proof:}(\underline{First part})  Let $1\leq r,s\leq n$ be fixed and $P_{m}=(i_{0}=r,i_{1},\ldots, i_{m-1},i_{m}=s)$ be given. We remove the vertex $i_1$
from the path $P_m$ to form the path $P_{m-1}=(i_{0}=r,i_{2},\ldots, i_{m-1},i_{m}=s)$. Then $P_m$ is a $m-$path from the vertex $r$ to the
vertex $s$ and $P_{m-1}$ is a $(m-1)$-path from the vertex $r$ to the vertex $s$. Then,
$$w(P_m)=a_{i_0i_1}*a_{i_1i_2}*\ldots*a_{i_{m-1}i_m}$$  and
$$w(P_{m-1})=a_{i_0i_2}*a_{i_2i_3}*\ldots*a_{i_{m-1}i_m}.$$
Then,
\begin{eqnarray*}
|w(P_{m})-w(P_{m-1})|&=&|(a_{i_0i_1}*a_{i_1i_2}*\ldots*a_{i_{m-2}i_{m-1}})*a_{i_{m-1}i_m}\\
&&-(a_{i_0i_2}*a_{i_2i_3}*\ldots*a_{i_{m-2}i_{m-1}})*a_{i_{m-1}i_m}|\\
&\leq&\alpha|(a_{i_0i_1}*a_{i_1i_2}*\ldots*a_{i_{m-3}i_{m-2}})*a_{i_{m-2}i_{m-1}}\\
&&-(a_{i_0i_2}*a_{i_2i_3}*\ldots*a_{i_{m-3}i_{m-2}})*a_{i_{m-2}i_{m-1}}|\\
&\vdots&\\
&\leq&\alpha^{m-2}\langle 1,0\rangle \hspace{0.1cm}(\mbox{by the repeated application of Equation (\ref{e6}})).
\end{eqnarray*}

This implies, with the help of Lemma \ref{L2}
\begin{eqnarray*}
w(P_m)\leq w(P_{m-1})+\alpha^{m-2}\langle 1,0\rangle\leq [A^{m-1}]_{rs}+\alpha^{m-2}\langle 1,0\rangle.
\end{eqnarray*}

Since $P_m$ is an arbitrary path of length $m$, this leads the following inequality hold
\begin{eqnarray}\label{e9}
[A^{m}]_{rs}\leq[A^{m-1}]_{rs}+\alpha^{m-2}\langle 1,0\rangle.
\end{eqnarray}

On the other hand, let $P_{m-1}=(i_{0}=r,i_{2},\ldots, i_{m-1},i_{m}=s)$ be a $(m-1)$-path from the vertex $r$ to the vertex $s$.
Choose $1\leq i_{1}\leq n$. Let $P_{m}=(i_{0}=r,i_{1},\ldots, i_{m-1},i_{m}=s)$, then $P_m$ is a $m-$path from the vertex $r$
to the vertex $s$. Then we have by the help of Equation (\ref{e6}),
\begin{eqnarray*}
w(P_{m-1})\leq w(P_m)+\alpha^{m-2}\langle 1,0\rangle\leq [A^k]_{rs}+\alpha^{m-2}\langle 1,0\rangle.
\end{eqnarray*}

This implies,
\begin{eqnarray}\label{e10}
 [A^{m-1}]_{rs}\leq [A^{m}]_{rs}+\alpha^{m-2}\langle 1,0\rangle.
\end{eqnarray}

From (\ref{e9}) and (\ref{e10}), we obtain $|[A^m]_{rs}-[A^{m-1}]_{rs}|\leq \alpha^{m-2}\langle 1,0\rangle$.

Let $N$ be a fixed natural number and for all $m\geq N$,
\begin{eqnarray*}
|[A^m]_{rs}-[A^N]_{rs}|&\leq& |[A^m]_{rs}-[A^{m-1}]_{rs}|+|[A^{m-1}]_{rs}-[A^{m-2}]_{rs}|+\ldots +|[A^{N+1}]_{rs}-[A^{N}]_{rs}| \\
&\leq& \alpha^{m-2}\langle 1,0\rangle+\alpha^{m-3}\langle 1,0\rangle+\ldots+\alpha^{N-1}\langle 1,0\rangle\\
&\leq& \alpha^{N-1}\Big(\frac{\langle 1,0\rangle}{1-\alpha}\Big).
\end{eqnarray*}

Since $0\leq \lambda\leq1$, we have the sequence $\{[A^{m}]_{rs}\}$ is a Cauchy sequence and hence convergent. That imply, $\lim \limits_{m\rightarrow \infty}A^{m}={\AA}$.

(\underline{Second part}) Let $P_{m}(rj)=(i_{0}=r,i_{1},\ldots, i_{m-1},i_{m}=j)$ be a $m$-path from the vertex $r$ to the vertex $j$ and
$P_{m}(sj)=(i_{0}=s,i_{1},\ldots, i_{m-1},i_{m}=j)$ be another $m$-path from the vertex $s$ to the vertex $j$. Then, by Equation (\ref{e6})
\begin{eqnarray*}
|w(P_m(rj))-w(P_m(sj))|\leq \alpha^{m-2}\langle 1,0\rangle.
\end{eqnarray*}
Now, by Lemma \ref{L2}
\begin{eqnarray*}
w(P_m(rj))&\leq& w(P_m(sj))+\alpha^{m-2}\langle 1,0\rangle\leq [A^m]_{sj}+\alpha^{m-2}\langle 1,0\rangle \hspace{0.1cm}\mbox{and}\\
w(P_m(sj))&\leq& w(P_m(rj))+\alpha^{m-2}\langle 1,0\rangle\leq [A^m]_{rj}+\alpha^{m-2}\langle 1,0\rangle.
\end{eqnarray*}
From the above two inequalities, $|[A^{m}]_{rj}-[A^{m}]_{sj}|\leq \alpha^{m-2}\langle 1,0\rangle $.

Now as, $\lim \limits_{m\rightarrow \infty}A^{m}={\AA}$, we can obtain ${\AA}_{rj}={\AA}_{sj}$.

\begin{theorem}\label{t7}
Let $A$ be an $n\times n$ IFM and $\lim \limits_{m\rightarrow \infty}A^{m}={\AA}$. Then all entries in the $j$-th column of
${\AA}$ are $\langle 1,0\rangle$, that is, $[{\AA}]_{sj}=\langle 1,0\rangle$ for all $s=1,2,\ldots, n$ if and only if there
is a critical path in $G$ from a critical vertex to the vertex $j$.
\end{theorem}
{\bf Proof:} This theorem can be proved by the same procedure as in Theorem \ref{t2} and with the help of Theorem \ref{t6}.

\begin{example}
Let us consider the IFM $B=$$\left[
   \begin{array}{ccc}
   \langle 0,1\rangle & \langle 1,0 \rangle & \langle 0.5,0.4 \rangle\\
   \langle 1,0 \rangle & \langle 0,1 \rangle & \langle 1,0 \rangle \\
   \langle 0.6,0.3 \rangle & \langle 1,0 \rangle & \langle 0,1 \rangle\\
   \end{array}
\right ]$.

Then the directed IFG corresponding to the IFM $B$ is given in Figure \ref{fig4}.
\begin{figure}[h]
\begin{center}
\unitlength 1mm 
\linethickness{0.4pt}
\ifx\plotpoint\undefined\newsavebox{\plotpoint}\fi 
\begin{picture}(54.75,44.375)(0,0)
\put(10.25,25.25){\circle{9.179}}
\put(47.839,25.839){\circle{9.179}}
\put(25.839,2.839){\circle{9.179}}
\put(29.5,37.25){\vector(1,0){.07}}\qbezier(9.25,29.75)(30.625,44.375)(47.5,30.5)
\put(28.25,22){\vector(-1,0){.07}}\qbezier(43.25,25.25)(27.75,19.25)(14.25,24.25)
\put(47.25,8){\vector(-3,-4){.07}}\qbezier(50.75,23.5)(54.75,4.25)(28.75,0)
\put(34.5,17){\vector(3,4){.07}}\qbezier(28.5,6.25)(32.75,19.25)(44,23.25)
\put(8.75,7.25){\vector(2,-3){.07}}\qbezier(6,23.25)(3.625,2.375)(21.75,1)
\put(21.5,17.5){\vector(-1,2){.07}}\qbezier(22.75,6)(25.125,20.875)(13,22.25)
\put(10.25,25){\makebox(0,0)[cc]{$v_1$}}
\put(47.5,25.5){\makebox(0,0)[cc]{$v_2$}}
\put(25.5,2.75){\makebox(0,0)[cc]{$v_3$}}
\put(27.5,40.5){\makebox(0,0)[cc]{$\langle 1,0\rangle$}}
\put(27,24.75){\makebox(0,0)[cc]{$\langle 1,0\rangle$}}
\put(49.5,4.75){\makebox(0,0)[cc]{$\langle 1,0\rangle$}}
\put(38.25,13.25){\makebox(0,0)[cc]{$\langle 1,0\rangle$}}
\put(-.5,8.5){\makebox(0,0)[cc]{$\langle 0.5,0.4\rangle$}}
\put(15.5,13.5){\makebox(0,0)[cc]{$\langle 0.6,0.3\rangle$}}
\end{picture}

\end{center}
\caption{Directed IFG $G$}\label{fig4}
\end{figure}
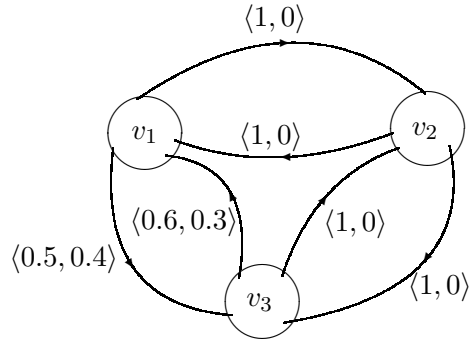
\end{example}
Here $v_1\rightarrow v_2\rightarrow v_1$ and $v_2\rightarrow v_3\rightarrow v_2$ are two different critical circuits.
Then the set of all critical vertices are $\{v_1,v_2,v_3\}$.

Now for $\lambda=0.5$, the eighth power of $B$ is,\\
$B^{8}=$$\left[
   \begin{array}{ccc}
   \langle 1,0 \rangle & \langle 0.93326,0.05339 \rangle & \langle 1,0 \rangle\\
   \langle 0.94661,0.04004 \rangle & \langle 1,0 \rangle & \langle 0.94661,0.04004 \rangle \\
   \langle 1,0 \rangle & \langle 0.94661,0.04004 \rangle & \langle 1,0 \rangle\\
   \end{array}
\right ]$ and \\
the limit matrix is, $\hat{B}=B^{28}=$$\left[
   \begin{array}{ccc}
   \langle 1,0 \rangle & \langle 1,0 \rangle & \langle 1,0 \rangle\\
   \langle 1,0 \rangle & \langle 1,0 \rangle & \langle 1,0 \rangle \\
   \langle 1,0 \rangle & \langle 1,0 \rangle & \langle 1,0 \rangle\\
   \end{array}
\right ]=U$.

\section{Conclusions}
Here we derive the procedure to get the power of an IFM under the maxgeneralized mean-mingeneralized mean
operation and the convex combination of max-min and maxarithmetic mean-minarithmetic mean operation using
the graph theoretic concept. In this paper, we showed that the power of an IFM with the said operations
are always convergent. Moreover, the limit IFM has the feature that all elements in each column are identical
for both operations defined above. In our further work, we shall try to test the convergence of IFMs with
respect to other binary operations.

\end{document}